\documentclass[12pt]{iopart}
\usepackage{graphicx}
\usepackage{harvard}
\usepackage{bbold}
\usepackage{blindtext}
\newcommand{\ket}[1]{|#1\rangle}             % distinguishabilty et Dirac's notation %
\newcommand{\bra}[1]{\langle#1|}             % Bra Dirac's notation  %
\begin{document}

\title{The duality principle in the presence of postselection}

\author{Jonathan Leach$^{1}$, Eliot Bolduc$^{1}$, Filippo.~M.~Miatto$^{2}$, Kevin Pich\'e$^{2}$, Gerd Leuchs$^{2,3}$, Robert W.~Boyd$^{2,4,5}$}
\address{$^1$IPaQS, SUPA, Heriot-Watt, Edinburgh, UK}
\address{$^2$Dept.~of Physics, University of Ottawa, 150 Louis Pasteur, Ottawa, Ontario, K1N 6N5 Canada}
\address{$^3$Max Planck Institute for the Science of Light, Erlangen, Germany}
\address{$^4$Institute of Optics, University of Rochester, Rochester, USA}
\address{$^5$SUPA, School of Physics \& Astronomy, University of Glasgow, G12 8QQ, UK}
\date{\today}

\ead{j.leach@hw.ac.uk}

\begin{abstract}
The duality principle, a cornerstone of quantum mechanics, limits the coexistence of wave and particle behaviours of quantum systems.  This limitation takes a quantitative form when applied to the visibility $\mathcal V$ and predictability $\mathcal P$ within a two-alternative system, which are bound by the inequality  $\mathcal  V^2+  \mathcal P^2 \leq 1$. However, if such a system is coupled to its environment, it becomes possible to obtain conditional measures of visibility and predictability, i.e.~measures that are conditional on the state of the environment. We show that in this case, the predictability and visibility values can lead to an apparent violation of the duality principle.  We experimentally realize this apparent violation in a controlled manner by enforcing a fair-sampling-like loophole via postselection.  Given the ability to simultaneously obtain high predictability information and high visibility interference fringes for a wide range of coupling strengths and postselected states, this work highlights the role of fair-sampling in tests of the duality principle.
\end{abstract}

\vspace{2pc}

\maketitle

%{\bf The double-slit experiment is the quintessential example of dualism, while the Mach-Zender interferometer is its archetypical model.  } 

\section{Introduction}
The duality principle provides us with one of the most well-known statements about quantum mechanics: the presence of interference and the existence of which-alternative information are mutually exclusive. For the case of such two-dimensional systems or qubits, Greenberger and Yasin defined a quantitative measure of which-alternative information, which we refer to as the predictability $\mathcal P$, and a quantitative measure of interference, the visibility $\mathcal V$ \cite{greenberger1988simultaneous}. They demonstrated the result $\mathcal  V^2+  \mathcal P^2 = 1$ for pure states. They also showed that this formula, in case of a qubit embedded in an environment, generalizes to the duality relation
\begin{equation}\label{VD}
\mathcal  V^2+  \mathcal P^2 \leq 1.
\end{equation}
Embedding the qubit in an environment enables realistic situations to be considered.   In this case, multiple degrees of freedom, each acting as a quantum system, can potentially be coupled together.
Englert studied the effect of coupling a qubit to an environment more formally in \cite{englert1996fringe}. The deep significance of the duality principle is in the fact that the quantities involved bound each other: the more is known about the alternatives, the less they can interfere and vice versa. This principle has been put to the test, directly and indirectly, many times and in different regimes \cite{aspect1987,durr1998fringe,schwindt1999quantitative,zeilingerC60,KolesovPlasmon,kocsis2011observing}. In all of the experimental tests, the duality principle prevailed.  However, while not in conflict with the duality principle, Menzel {\it et al.}~recently reported high which-alternative information and high-visibility fringes in a single experiment \cite{menzel2012wave,bolduc2014menzel}. 

Bergou and Englert have shown various duality relations that apply in the case of a qubit coupled to an arbitrarily large environment \cite{englert2000quantitative}. In particular, the most stringent one is 
\begin{eqnarray} \label{eq:bergou}
\mathcal{V}_{\hat \pi}^2+\mathcal{P}_{\hat \pi}^2\leq1,
\end{eqnarray}
 where $\hat \pi$ is an observable of the environment \footnote{The notation $\mathcal{V}_{\hat \pi}$ should be read as the visibility of the conditional state $\hat {\rho}$, obtained after a successful postselection of the observable $\hat {\pi}$.  It is shorthand notation for $\mathcal{V}(\hat{\rho}| \hat \pi)$.}.   It follows from this equation that for every state of the environment, the duality principle prohibits simultaneous knowledge of visibility and predictability.

%{\bf In this work, we  demonstrate experimentally that when the visibility and the predictability, $\mathcal V$ and $\mathcal P$, are measured with respect to different environment outcomes, 

In this work, we demonstrate experimentally the conditions in which it is possible to obtain both high-visibility interference fringes and high which-way information in a single experiment.  At first, this result may seem in conflict with the principles of duality; however, it can be explained simply using the concepts of coupling, postselection, and unfair sampling. The key feature that enables unfair sampling is a non-separable state of the system of interest and its environment.  Such correlation allows the measurements of visibility and predictability to be conditioned on successful postselection of different environment outcomes, say $\hat \pi_1$ and $\hat \pi_2$, and consequently measure high values for each simultaneously.  

The explicit manner in which we control the coupling between different degrees of freedom may seem extreme.  However, we note that coupling between degrees of freedom occurs naturally in many physical processes, and it is the key concept for generalised measurements.  In addition, postselection of a distribution occurs in many experiments, e.g., when single-mode fibres are used to collect the fundamental mode of a field.  The control of coupling and postselection therefore highlights one potential method that can lead to an unfair sampling a system.  We find that dramatic results occur for a range of coupling strengths and find that even only weak coupling results in an apparent violation.  As only weak coupling is needed, this work serves to guide future experiments on tests of the duality principle.

An interesting outcome of this work is the analogy between conditional measurements of visibility associated with the duality principle and the class of weak measurements known as ``direct measurements'' \cite{salvail2013full,lundeen2011direct}. The analogy is due to a symmetry of the measurement procedures: the roles of system and environment are interchanged in the two cases. In the case of conditional visibility, a measurement on the system is performed after successful selection of a state of the environment, while in the case of weak measurement, a measurement on the environment (usually called ``pointer'') is performed after successful selection of a state of the system \cite{aharonov1988result,ritchie1991realization}. The analogy between these two procedures indicates that one can understand our experimental findings within the framework of weak measurement. %This result provides insight into the nature of weak measurement and is consistent with the recent notion that weak values are present in all von Neumann measurement schemes \cite{dressel2012weak}.
 
%%%%%%%%
% THEORY
%%%%%%%%

\section{Theory}

We illustrate the subtleties of fair sampling applied to the duality principle by considering a simple example where two internal degrees of freedom of a single physical system are coupled together and act as qubit and environment. We first consider the predictability and visibility measures of the qubit when the environment plays no role; the results are consistent with our conventional understanding of the duality principle.  We then go on to show that an apparent violation of Eq.~(\ref{VD}) can be obtained when conditional measurements are performed.
% The two degrees of freedom are used to represent the qubit and the environment. 

To keep the treatment simple, and without losing any power in our arguments \footnote{thanks to the possibility of employing a Schmidt decomposition on the joint state}, we consider an environment that is also a qubit. A convenient way of realizing this situation in an optics framework is by using two eigenmodes of orbital angular momentum (OAM) of value $+\ell$ and $-\ell$, which we use as the qubit, and the polarization degree of freedom, which we use as the environment.
 
In our setup we produce the following state of OAM and polarization: 
\begin{equation}\label{psi}
%\ket{\Psi}=  &c_\theta \ket{\ell}\ket{V} + s_\theta \ket{-\ell} \left(c_\alpha \ket{H} +  s_\alpha \ket{V} \right),\\
\ket{\Psi}=   \cos \left(\frac{\theta}{2} \right) \ket{\ell}\ket{V} + \sin \left(\frac{\theta}{2} \right) \ket{-\ell} \left( \cos \left(\frac{\alpha}{2} \right) \ket{H} +   \sin \left(\frac{\alpha}{2} \right) \ket{V} \right).  \nonumber
\end{equation}
%
%where $c_\theta=\cos(\theta/2)$ and $s_\theta=\sin(\theta/2)$ and similarly for $\alpha$.  
This state can be either separable or nonseparable; the degree of nonseparability can be controlled by the two angles $\theta$ and $\alpha$.  In particular, the state is nonseparable and maximally correlated for the configuration $\theta=\pi/2$ and $\alpha=0$.
The density operator of the combined OAM and polarizaton qubits is given by $\hat\Psi = \ket{\Psi}\bra{\Psi}$. Therefore, the  state of the OAM qubit ignoring polarization is given by the partial trace:
\begin{equation}\label{partialTr}
\hat\rho_\ell = {\rm Tr}_\mathrm{pol}[\hat\Psi].
\end{equation}
The visibility and predictability associated with the OAM qubit $\hat\rho_\ell$ can be expressed in a compact way by using the Pauli operators in the OAM space applied in the following way \cite{englert1996fringe}: $\mathcal V =  |{\rm Tr}[ (\hat\sigma_x+i\hat\sigma_y) \hat \rho_\ell ]| = |\cos (\alpha/2) \sin {\theta}|$ and $\mathcal P =  |{\rm Tr}[\hat \sigma_z \hat \rho_\ell ] |= | \cos{\theta}|$. Taking the sum of the squares of these quantities, we find 
\begin{equation}\label{aveVDenv}
%\mathcal V^2 + \mathcal P^2 = s_\alpha^2s_{2\theta}^2+c_{2\theta}^2\leq 1.\\
\mathcal V^2 + \mathcal P^2 = \sin^2 \left(\frac{\alpha}{2} \right) \sin^2{\theta}+\cos^2{\theta}\leq 1.
\end{equation}
One can see that  no values of $\alpha$ and $\theta$ lead to a violation of the duality principle, which is consistent with the duality relation of Eq.~\ref{VD}.

Now consider exploiting the correlation between the two qubits to obtain \emph{conditional} values of $\mathcal V$ and $\mathcal P$. This idea has been previously explored theoretically by Bergou and Englert \cite{englert2000quantitative}.
The state of the OAM qubit conditioned on successful postselection of the polarization degree of freedom is
\begin{equation}
\hat\rho_{\ell|\hat\pi} = \frac{{\rm Tr}_\mathrm{pol}[(\hat{\mathbb{1}}\otimes\hat\pi)\hat\Psi]}{p}.
\end{equation}
Here, $\hat {\pi}$ is a state of polarization, $p=\mathrm{Tr}[ (\hat{\mathbb{1}} \otimes \hat \pi) \hat\Psi]$ is the postselection probability, and the vertical bar notation means ``given successful postselection on''. The visibility and the predictability of the OAM qubit that are conditioned on a successful postselection of the polarization qubit are
\begin{eqnarray}
\mathcal{V}_{\hat \pi} =|\mathrm{Tr}[ (\hat\sigma_x+i\hat\sigma_y)\hat\rho_{\ell| \hat \pi}]|
\quad \mathrm{and } \quad 
\mathcal{P}_{\hat \pi} =|\mathrm{Tr}[ \hat\sigma_z \hat\rho_{\ell|\hat\pi}]|.\label{CondValb}
\end{eqnarray}
These two quantities will satisfy the duality relation, Eq.~\ref{eq:bergou}, if the postselection used in both cases is the same.   However, seemingly contradictory results can be obtained when measuring visibility and predictability conditioned on \emph{different} postselections $\hat\pi_1$ and $\hat\pi_2$. The most extreme case is when $\hat\pi_1$ and $\hat\pi_2$ are orthogonal to each other.  In this case one obtains
\begin{equation}
\mathcal{V}_{\hat \pi_1}^2+ \mathcal{P}_{\hat \pi_2}^2 \leq 2,
\label{ineq2}
\end{equation}
as $\mathcal{V}_{\hat \pi_1}$ and $ \mathcal{P}_{\hat \pi_2} $ can independently reach 1.  % {\bf One could, of course, interchange the roles of $\hat {\pi_1}$ and $\hat {\pi_2}$, using $\mathcal{V}_{\hat \pi_2}$ and $\mathcal{P}_{\hat \pi_1}$ instead, and obtain an equivalent result.}

In our experiment, the visibility and predictability that are obtained by postselecting the state $|\Psi\rangle$ on vertical and horizontal polarizations (i.e. $\hat\pi_V=|V\rangle\langle V|$ and  $\hat\pi_H=|H\rangle\langle H|$) are
\begin{eqnarray}
\label{Vcond}
%\mathcal V(\hat\pi_V)&=\frac{|s_{2\theta}s_\alpha|}{c_\theta^2+s_\theta^2s_\alpha^2}\\
\mathcal{V}_{\hat \pi_V} =\frac{| \sin{\theta} \sin (\frac{\alpha}{2})|}{\cos^2 (\frac{\theta}{2})+\sin^2 (\frac{\theta}{2}) \sin^2 (\frac{\alpha}{2})} \hspace{10pt}  \quad \mathrm{and } \quad 
\mathcal{P}_{\hat \pi_H} =1. \label{Pcond}
\end{eqnarray}
 The sum of the squares of $\mathcal{V}_{\hat \pi_V}$ and $\mathcal{P}_{\hat \pi_H}$ is always bounded below by 1 and above by 2.  

This result is in apparent violation of the duality principle since visibility and predictability can both reach 1 for a given configuration of $\alpha$ and $\theta$.   The origin of this outcome is one of the main results of this paper: the apparent violation can only occur if the postselected states differ from each other.  The orthogonality between our postselections, $\hat {\pi}_H$ and $\hat {\pi}_V$,  is a deliberate choice that produces the extreme results, but an apparent violation can occur for other choices as well.    From inspection of $\ket{\Psi}$ (Eq.~\ref{psi}), we see that $\ket {-\ell}$ is the only OAM component associated with the horizontal degree of freedom.  Therefore, a measurement of predictability conditioned on successful postselection of the horizontal polarization state will always equal to 1.  Consider that this result combined with any measure of visibility that is greater than zero will result in an apparent violation of the duality principle.   From the construction of our state, the only postselected polarization that will have zero visibility is the horizontal state; however, this is not the case if we consider the vertical postselection.  In this case, it is possible to achieve high-visibility fringes, albeit with an associated low probability.  Combining the two results, each obtained with different postselections, gives rise to a sum of the squares of the visibility and predictability measures that is greater than 1.    

%By taking all of the postselected outcomes of polarization into account, one can calculate the average of the square of the visibilities and predictabilities as defined by 
%%
%\begin{eqnarray}
%\langle\mathcal V^2\rangle&=\sum_{k=H,V} p_k\mathcal V(\hat\rho_{\ell|\hat\pi_k})^2 = \sum_{k=H,V} p_k\mathcal V_{| \hat\pi_k}^2  \label{V2}\\
%\langle\mathcal P^2\rangle&=\sum_{k=H,V} p_k\mathcal P(\hat\rho_{\ell|\hat\pi_k})^2 = \sum_{k=H,V} p_k\mathcal P_{| \hat\pi_k}^2  \label, \label{P2}
%\end{eqnarray}
%%
%%
%where $p_k$ is the probabilities of occurrence of the $k^{\rm{th}}$ outcome. In the case of a pure state, we have $\mathcal V(\hat\rho_{\ell|\hat\pi_k})^2+\mathcal P(\hat\rho_{\ell|\hat\pi_k})^2=1$ for each polarization outcome $\hat{\pi}_k$, and the sum of Eq.~\ref{V2} and Eq.~\ref{P2} is thus also equal to one. The equality $\langle\mathcal V^2\rangle^2+\langle\mathcal P^2\rangle^2=1$ is in principle satisfied invariantly of the initial state. {\it One can extremely easily check that the equality is satisfied for any value of $\alpha$ and $\theta$; if you can't, you should seriously reconsider your career choice.}

One way to interpret this result is to associate a different qubit state for every state of the environment, i.e., the OAM qubit associated with horizontal polarization is different to that associated with vertical polarization. It is the postselection that alters the qubit.  Consequently, one can measure visibility and predictability of each postselected qubit independently of the other.   When the visibility and predictability of the OAM qubit vary from one state of the environment to the next, it is then easy to postselect a state with either high visibility or high predictability.   In other words, states with high visibility and states with high predictability are both available.  

The source of contradiction is specifically postselection: when there are many conditional measurements available, each with an associated probability of occurrence, the duality principle is satisfied if applied to the averaged visibility and predictability: 
\begin{eqnarray}
\overline{\mathcal V} = \sum_k p_k\mathcal V_{\hat\pi_k} \quad \mathrm{and } \quad \label{eq:avgV} 
\overline{\mathcal P} =\sum_k p_k\mathcal P_{\hat\pi_k}. \label{eq:avgP}
\end{eqnarray}
where in our case $k$ labels vertical and horizontal polarizations.
For the state $|\Psi\rangle$, we have $\overline{\mathcal V}  =| \sin (\theta) \sin (\alpha/2)|$ and $ \overline{\mathcal P} = \sin^2 (\theta/2) \cos^2(\alpha/2)+|\cos^2 (\theta /2)-\sin^2 (\theta /2)\sin^2 (\alpha/2))|$, the sum of the squares of which is always bounded by 1.  We see here that when  postselection probabilities are considered in addition to the outcomes of the conditional measurements, there is no violation of the duality principle.   This result applies to any combined system and environment \cite{englert2000quantitative}. 

\section{Experiment}
\begin{figure*}
\begin{center}
\includegraphics[width=0.9\textwidth]{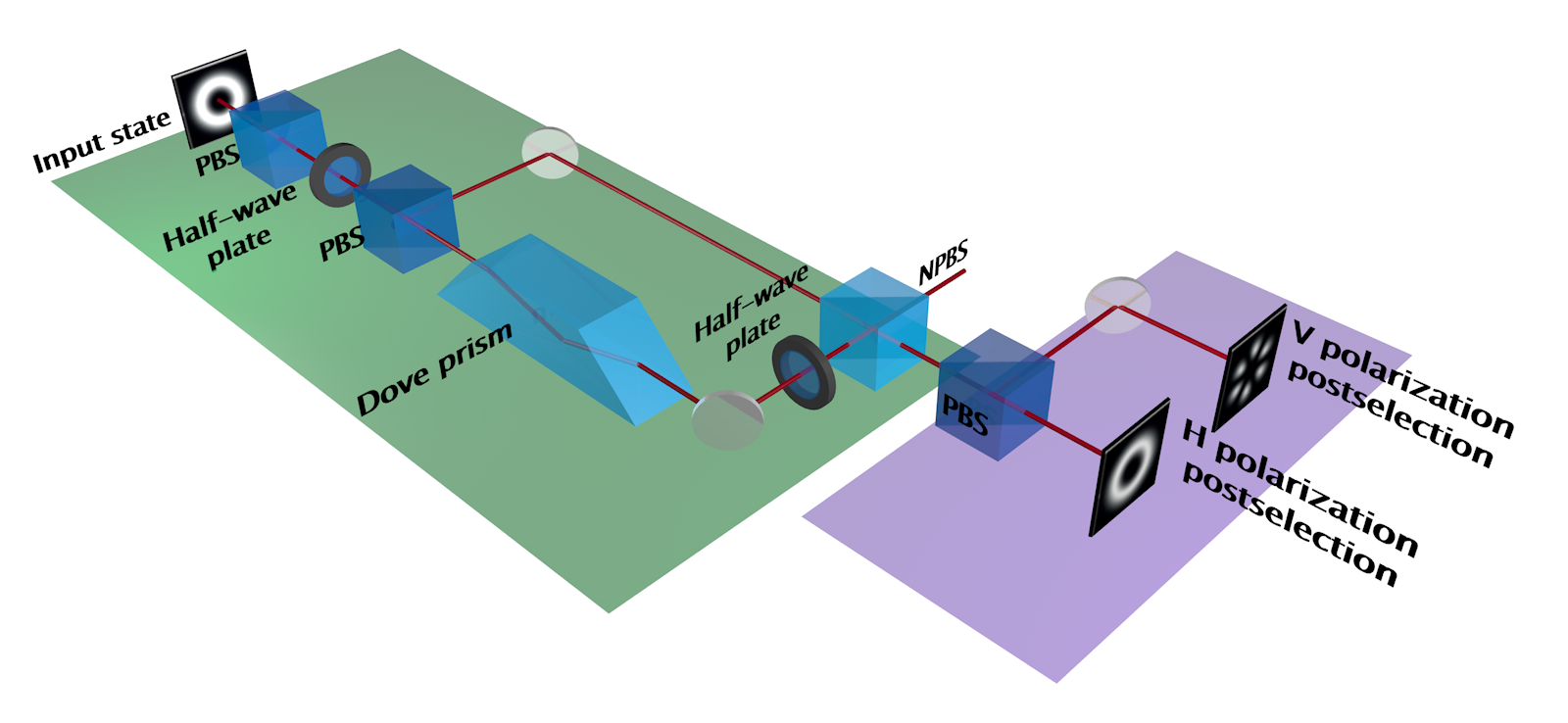}\label{exp}
\caption{We first prepare a non-separable state of OAM and polarization; this is indicated by the green area.  We then perform postselection; this is indicated by the purple area.  The state of OAM is generated using a HeNe laser and a spatial light modulator (not shown).   We control the amplitude in each path of the interferometer with a half-wave plate and a polarizing beam splitter (PBS) (this controls $\theta$).  Inside the lower path, a Dove prism reverses the handedness of the OAM mode and a second half-wave plate controls the polarization state inside one arm of the interferometer (this controls $\alpha$).  The non-polarizing beam splitter (NPBS) produces a superposition of the two paths, and the final PBS allows postselection on polarization. We measure the conditional visibilities and predictabilities for the $V$ and the $H$ outputs using images captured with a CCD camera.}
\label{default}
\end{center}
\end{figure*}
\begin{figure*}
\begin{center}
\includegraphics[width=1\textwidth]{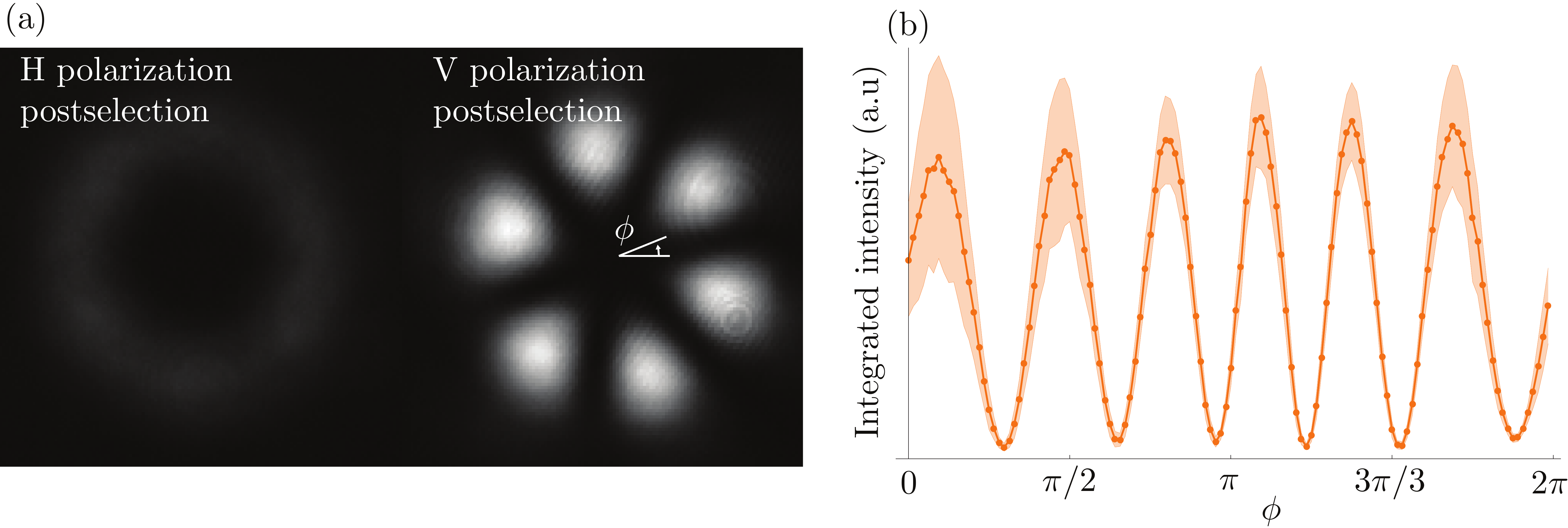}\label{typicalimage}
\caption{(a) A typical image of the $V$ and the $H$ posteselected outputs  captured with the CCD camera.  The lefthand section of the image, the horizontal postselection, show a very faint $\ell = -3$ mode; there is high predictability  $\mathcal P_{\hat\pi_H}= 0.98$ associated with this outcome.  The righthand section of the image, the vertical postselection, shows the intensity of a superposition of $\ell = -3$ and $\ell = +3$ modes; there are high visibility fringes  $\mathcal V_{\hat\pi_V} = 0.93$ associated with this outcome.  The value of  $\mathcal{V}_{\hat \pi_V}^2+ \mathcal{P}_{\hat \pi_H}^2$ is equal to 1.83. (b)  The azimuthally integrated intensity of the vertical postselection shown in (a).   Each data point corresponds to the average intensity in a 3$^\circ$ angular window.  The shaded region indicates the error band, which is at one $\sigma$. }
\label{default}
\end{center}
\end{figure*}
The goal of the experiment is to perform the conditional measurements outlined in the previous section, highlighting the significance of coupling and postselection in tests of the duality principle. We first prepare a non-separable state of OAM and polarization; we then perform postselection of the polarization degree of freedom; and finally, we calculate the conditional visibility and predictability measures of the OAM degree of freedom.

The state of the system and environment is generated by inserting an OAM mode of $\ket{\ell = +3}$ into a Mach-Zehnder interferometer with a Dove prism and half-wave plate in one of the arms; see figure 1.   Before the interferometer, we use a collimated HeNe and a spatial light modulator to generate the OAM mode.  The interferometer performs the role of ``entangling" the OAM and polarization degrees of freedom, resulting in a non-separable state.  The precise form of the joint OAM-polarization state is controlled by the two half-wave plates. The half-wave plate before the interferometer controls $\theta$ in the state; the half-wave plate inside the interferometer controls $\alpha$.

A polarizing beam splitter after the output port of the interferometer projects onto the horizontal and vertical states of polarization.   The horizontal polarization output is always composed of a single OAM mode, while the vertical polarization output is generally composed of two OAM modes of opposite handedness and with varying amplitudes, leading to a petal-shaped interference pattern.

We use a CCD camera to record intensities of the modes and then calculate the visibility and distinguishability measures. Figure 2 shows a typical image captured by the camera. In order to obtain an apparent violation of the duality relation in the way described above, we measure the predictability after postselection of horizontal polarization and the visibility after postselection of vertical polarization. We measure $\mathcal P_{\hat\pi_H}$ as the difference in intensity of the two arms of the interferometer. Then we measure the visibility $\mathcal V_{\hat\pi_V}$ with respect to the vertical polarization.  This is calculated by integrating radially with respect to the centre to obtain a plot of intensity versus angle and measuring the visibility of the curve that is obtained.

\section{Results}

\begin{figure}[t!]
\begin{center}
\includegraphics[width=0.9\columnwidth]{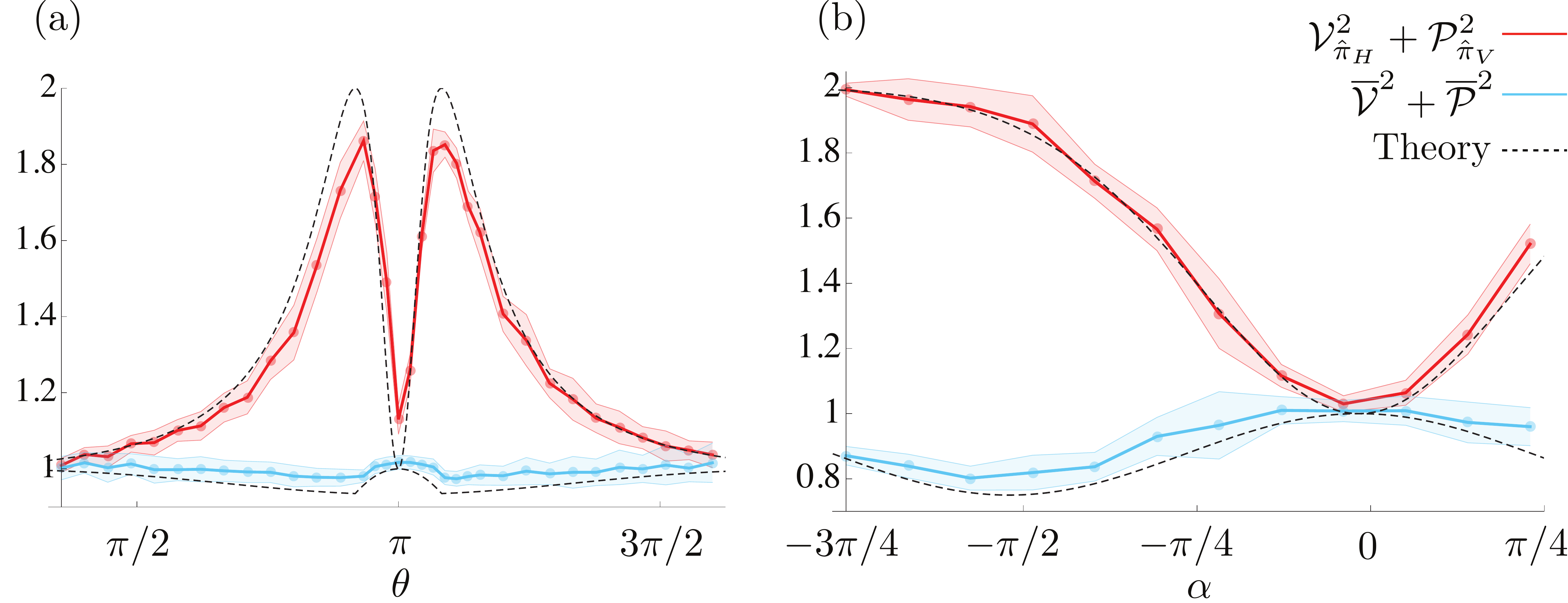}
\caption{\label{tan} (a) Duality relation as a function of the initial polarization for a weak environment coupling $\alpha=\pi/12$. (b)  Duality relation as a function of coupling for a fixed initial polarization $\theta = \pi/2$. The red lines show that it is possible obtain values of $\mathcal V_{\hat {\pi}_H}^2 +\mathcal P_{\hat {\pi}_V}^2$ higher than 1 for a large range of states and couplings.  The blue lines show that when the average quantities are used, the sum is never greater than 1. For the plots given above, the error band is at one $\sigma$.}
\label{default}
\end{center}
\end{figure}

Our experimental results are shown in figure \ref{tan}(a) and (b).   The data show the sum of the squares of the measured conditional visibilities and predictabilities together with the average quantities.   Figure \ref{tan}(a) shows data for a range of values of $\theta$; recall that this controls the polarization state before the interferometer.   Figure \ref{tan}(b) shows data for a range of values of $\alpha$; this controls the polarization state of the lower path of the interferometer.  In both figures, we see that the sum of the squares of the conditional measurements exceeds 1 (see the red curves).  In contrast, the sum of the squares of the averaged quantities never exceeds 1 (see the blue curves).   

Consider the result in figure \ref{tan} (a), where the highest value of $\mathcal V_{\hat\pi_V}^2 + \mathcal P_{\hat\pi_H}^2$ appears when $\theta = \pi \pm \alpha$, where $\alpha=\pi/12$ in the example is a small angle.  This state corresponds to when the input polarization state is close to, but not quite, horizontal.  In this case, almost all the light enters the lower arm of the interferometer, but due to the small component of vertical polarization of the input state, there will be a small component in the upper arm.   The small amplitude of the vertical polarization state in the upper arm can be matched in the lower arm by rotating the wave plate ($\alpha$).    The light that exits the interferometer now has a large horizontal component that only passed through the lower arm and a small vertical component that has passed through both arms.  A measurement of the predictability conditioned on the horizontal polarization state will be equal to unity, and a measurement of the visibility conditioned on the vertical polarization state will also be equal to unity.  Under these conditions, we can claim that we observe both high visibility and high predictability in a single experiment without violating the principle of duality.

%%%%%%%%%%%%%%%%%%
%   A USEFUL ANALOGY
%%%%%%%%%%%%%%%%%%
\section{{\bf The analogy with direct measurement}}

There are three main steps required for this work: coupling of the OAM and polarization degrees of freedom to produce a non-separable state, postselection of the polarization qubit, and measurement of properties of the OAM qubit.   These three concepts - coupling, postselection and measurement - are precisely those that are used in the procedure necessary to obtain weak values.  Consequently, there is a helpful analogy between the work that we present here and the recent work of ``direct measurement" \cite{lundeen2011direct,salvail2013full} that relies on weak values.  

We first summarize the result in \cite{lundeen2011direct}.  The physical framework of the reference is that of a two-alternative system. The two alternatives therein are for light to pass through or not to pass through a small half-wave plate sliver, positioned at transverse coordinate $x$. In case a photon passes through the sliver, its polarization is linearly rotated by a small angle $\varphi$. How small $\varphi$ needs to be in order to apply the following steps depends on the magnitude of the amplitudes corresponding to the two paths, as we will see. The unitary operator that produces such transformation, linearized to first order is $\hat U=\mathbb{ 1}-i\frac{\varphi}{2}\hat\sigma_y\otimes|x\rangle\langle x|$. Starting with the state $|V,\psi\rangle$, where $|\psi\rangle$ indicates a transverse profile, and applying $\hat U$, one obtains
\begin{equation}
\label{jeff1}
|V,\psi\rangle+\frac{\varphi}{2}|H\rangle \hat\pi_x|\psi\rangle,
\end{equation}
where $\hat\pi_x=|x\rangle\langle x|$. Notice that this state is not yet normalized. Finally postselecting on an unbiased state (in the reference, the state had a well defined transverse momentum $p=0$) results in a final state of polarization $|s\rangle=|V\rangle+\frac{\varphi}{2}\frac{\psi(x)}{\tilde\psi(0)}|H\rangle$, which we can consider ``normalized'' under the weakness condition $\left|\frac{\varphi}{2}\frac{\psi(x)}{\tilde\psi(0)}\right|\ll1$. The final step is to measure the expectation values of $\hat\sigma_x$ and $i \hat\sigma_y$, to reconstruct the weak value via the relation
\begin{equation}
\label{jeff}
\langle \hat\pi_x\rangle=\frac{\psi(x)}{\tilde\psi(0)}=\frac{1}{\varphi}\langle s|\hat\sigma_x-i\hat\sigma_y|s\rangle.
\end{equation}
This equation is the crucial ingredient of the analogy: it shows that the absolute value of the weak value multiplied by $\varphi$ is equal to the polarization visibility of the state $|s\rangle$.

The difference between our procedure and  ``direct measurement" is that we interchange system and environment: we measure the visibility of OAM (the system) after a postselection on the polarization (the environment), whereas, in direct measurement, one measures the visibility of the polarization (the environment or pointer) after a postselection on the position (the system). While the end goal of each method is different, the final step is to measure the visibility of a postselected state.  In the direct measurement method, one measures the visibility of the polarization in order to find the weak value, and in the conditional visibility method, one is interested in the visibility itself.

%%%%%%%%%%%%%%%%%%
%   CONCLUSIONS
%%%%%%%%%%%%%%%%%%

\section{Conclusions}

In this work, we show that if a qubit is coupled to its environment, it becomes possible to obtain simultaneous high values for conditional measures of visibility and predictability. In this case, where conditional measurements are made, the predictability and visibility values can lead to an apparent violation of the duality principle.   To achieve such a result, we are required to disregard certain measurement outcomes, enforcing a fair-sampling-like loophole via postselection.  However, we note that although our experimental procedure allowed us to purposely obtain simultaneous high values, there are realistic experimental cases where inadvertent postselection could be made.  One such example is a measurement that uses a single-mode fibre to capture the fundamental mode of a field.   In this case, care should be taken to ensure that the post-selected single mode is not coupled to the degree of freedom of interest.   We note that under no circumstance do we claim that a violation of the duality principle is possible; rather we seek to highlight certain experimental conditions where apparent violations occur.  As such, this work demonstrates the role of fair-sampling in tests of the duality principle and serves as a guide for future experiments.

\ack
This work was supported by the Canada Excellence Research Chairs (CERC) Program and the Natural Sciences and Engineering Research Council of Canada (NSERC).

%\end{harvard}

\end{document}